\RequirePackage{fix-cm}
\documentclass[11pt]{scrartcl}

\usepackage[table]{xcolor}
\usepackage{amssymb,amsmath,amsthm}
\usepackage{graphicx}
\usepackage{todonotes}
\usepackage{authblk}
\usepackage{color}
\usepackage{algorithm}
\usepackage[noend]{algpseudocode}
\algnewcommand\algorithmicinput{\textbf{Input:}}
\algnewcommand\algorithmicoutput{\textbf{Output:}}
\algnewcommand\Input{\item[\algorithmicinput]}%
\algnewcommand\Output{\item[\algorithmicoutput]}%
\usepackage{booktabs}
\usepackage{tabularx}
\usetikzlibrary{shapes}
\usepackage{url}
\usepackage{float}
\setcapindent{0pt}
\usepackage{physics}

\usetikzlibrary{arrows}
\usetikzlibrary{decorations.pathmorphing}

\usepackage[hidelinks,draft=false]{hyperref}
\usepackage[capitalise,nameinlink,noabbrev]{cleveref}

\newtheorem{theorem}{Theorem}[section]
\newtheorem{lemma}[theorem]{Lemma}
\newtheorem{corollary}[theorem]{Corollary}
\newtheorem{proposition}[theorem]{Proposition}
\newtheorem{observation}[theorem]{Observation}
\newtheorem{definition}[theorem]{Definition}

\renewcommand{\O}{\ensuremath{\mathcal{O}}}

\newcommand{\G}{\ensuremath{\mathcal{G}}}

\tikzstyle{vertex}=[inner sep=2pt,draw,circle, minimum size=8pt]
\tikzstyle{setedge} = [draw,line width=3]

\usepackage{amssymb,amsmath,amsthm}
\usepackage{graphicx,enumerate}
\usepackage{etoolbox,cite}
\usepackage{subfiles}
\usepackage{enumitem}

\usepackage{array}
\newcolumntype{x}[1]{>{\centering\arraybackslash\hspace{0pt}}p{#1}}

\usepackage{tikz}

\usepackage{hhline}
\usepackage{makecell}
\usepackage{multirow}
\usepackage{graphicx,enumerate}
\usepackage{etoolbox}
\usepackage{centernot}
\usepackage{algorithm}
\usepackage[noend]{algpseudocode}

\usetikzlibrary{decorations.pathmorphing}

\newcommand{\bf}{\bfseries}

\tikzstyle{vertex}=[inner sep=2pt,draw,circle, minimum size=8pt]
\tikzstyle{setedge} = [draw,line width=3]

\newcounter{caseCount}
\AtBeginEnvironment{cproof}{\setcounter{caseCount}{0}}

\newcounter{subcaseCount}[caseCount]

\newcounter{subsubcCount}[subcaseCount]

\usepackage{fullpage}

\begin{document}
\title{Sandwich Monotonicity and Recognition of Weighted Graph Classes\thanks{A preliminary version of parts of this work appeared in the proceedings of the 46th International Workshop on Graph-Theoretic Concepts in Computer Science (WG 2020)~\cite{MR4184212}. 
This research was funded in part by the German Academic Exchange Service and the Slovenian Research Agency (BI-DE/17-19-18 and BI-DE/19-20-007), and by the Slovenian Research and Innovation Agency  (I0-0035, research programs P1-0285, P1-0383, P1-0404 and research projects J1-60012, J1-70035, J1-70046, and N1-0370).}}
	\author[1]{Jesse Beisegel}
	\author[2,3]{Nina Chiarelli}
	\author[1]{Ekkehard K\"ohler}
	\author[2,3]{Matja\v{z} Krnc}
	\author[2,3]{\\Martin Milani\v c}
 	\author[3]{Nevena Piva\v c}
	\author[1]{Robert Scheffler}
	\author[4]{Martin Strehler}

	\affil[1]{\normalsize Brandenburg University of Technology, Institute of Mathematics, Cottbus, Germany}
	\affil[2]{\normalsize
  University of Primorska, FAMNIT, Koper, Slovenia}
	\affil[3]{\normalsize
  University of Primorska, IAM, Koper, Slovenia}
    \affil[4]{\normalsize Westsächsische Hochschule Zwickau, Department of Mathematics, Zwickau, Germany}

	\date{}

\maketitle

\begin{abstract}
Edge-weighted graphs play an important role in the theory of Robinsonian matrices and similarity theory, particularly via the concept of level graphs, that is, graphs obtained from an edge-weighted graph by removing all sufficiently light edges. This suggests a natural way of associating to any class $\G$ of unweighted graphs a corresponding class of edge-weighted graphs, namely by requiring that all level graphs belong to $\G$.
We show that for weighted graphs $G=(V,E)$ with weights from $\{1,\dots,|E|\}$ we can decide in linear time whether all level graphs are split, threshold, or chain graphs using special edge elimination orderings. 
We obtain these results by introducing the notion of degree sandwich monotone graph classes.
A graph class $\G$ is sandwich monotone if every edge set which may be removed from a graph in $\G$ without leaving the class also contains a single edge that can be safely removed. 
Furthermore, if we require the safe edge to fulfill a certain degree property, then $\G$ is called degree sandwich monotone. We present necessary and sufficient conditions for the existence of a linear-time recognition algorithm for any weighted graph class whose corresponding unweighted class is degree sandwich monotone and contains all edgeless graphs.
\end{abstract}

\section{Introduction}

\paragraph{Background.} Vertex and edge elimination orderings are well established concepts in graph theory~(see Chapter~5 in~\cite{brandstadt1999graph}). For example, chordal graphs can be characterized as the graphs with a perfect vertex elimination ordering~\cite{fulkerson1965incidence,rose1970triangulated}.
The vertices of a chordal graph can be deleted one by one in such a way that every vertex is simplicial at the time of removal; in particular, all the resulting graphs are chordal.

In 2017 Laurent and Tanigawa~\cite{laurent2017perfect} extended the classical notion of perfect (vertex) elimination ordering for graphs to edge-weighted graphs, giving a framework capturing common vertex elimination orderings of families of chordal graphs, Robinsonian matrices, and ultrametrics. 
They showed that an edge-weighted graph $G$ has a perfect elimination ordering if and only if it has a vertex ordering that is a simultaneous perfect elimination ordering of all its level graphs, where a \emph{level graph} of $G$ is any graph obtained from $G$ by removing all edges of weight below a certain threshold (we postpone the precise definition to \cref{sec:definition}). 
In particular, this latter condition implies that all the level graphs must be chordal.

Similarly, edge elimination orderings can be used to characterize graph classes. Adding an edge between a two-pair of a weakly chordal graph always maintains weak chordality~\cite{spinrad1995algorithms}. Since the class of weakly chordal graphs is self-complementary, a graph is weakly chordal if and only if it has an edge elimination ordering where every edge is a two-pair in the complement of the current graph. On a bipartite graph, an edge elimination ordering is said to be \emph{perfect} (also called perfect edge-without-vertex-elimination ordering) if every edge is bisimplicial at the time of elimination, that is, the closed neighborhood of both endpoints of the edge induces a complete bipartite subgraph. A bipartite graph is chordal bipartite if and only if it admits a perfect edge elimination ordering (see, e.g.,~\cite{MR1344547}).

What the above concepts have in common is that a certain property of the graph is maintained even after a certain vertex or edge has been deleted. Some graph properties achieve this in a trivial way. A graph class $\G$ is said to be  \emph{monotone} if every subgraph of a graph in $\G$ is also in $\G$. For example, planar graphs and bipartite graphs are monotone graph classes. In particular, an arbitrary edge can be deleted without leaving a monotone graph class.
Instead of deleting edges one at a time, one can also consider the deletion of pairwise disjoint sets of edges, sequentially.
All the edges from the current set are deleted at once and we require that all the intermediate graphs belong to a fixed graph class $\G$. 
An example of this process is given by the class of threshold graphs (see~\cite{chvatal1977aggregation}), where several edges may be deleted at once when the threshold for adjacency is raised. Furthermore, the level graphs of an edge-weighted graph, as used by Laurent and Tanigawa~\cite{laurent2017perfect}, are equivalent to such edge set eliminations.

These more general edge elimination sequences are naturally related to the following concept also studied in the literature. A graph class $\G$ is said to be \emph{sandwich monotone} if for any two graphs $G$ and $G'$ in $\G$ such that $G$ is a spanning subgraph of $G'$, the graph $G$ can be obtained from $G'$ by a sequence of edge deletions such that all the intermediate graphs are in $\G$. This property was studied in 1976 by Rose et al.~\cite{rose1976algorithmic}, who established it for the class of chordal graphs.
The term \emph{sandwich monotonicity} was introduced in 2007 by Heggernes and Papadopoulos~\cite{heggernes2007single} (see also~\cite{heggernes2009single}), who showed that the classes of threshold graphs and chain graphs are sandwich monotone. 
The same was shown for the class of split graphs by Heggernes and Mancini~\cite{heggernes2009minimal} as well as for classes of strongly chordal graphs and chordal bipartite graphs by Heggernes et al.~\cite{heggernes2011strongly}.

\paragraph{Our Contributions.}
To any class of unweighted graphs we associate a corresponding class of edge-weighted graphs. 
We introduce a novel approach to efficiently recognize some of these weighted graph classes using special edge elimination schemes.

Given a graph class $\G$, we say that an edge-weighted graph is \emph{level-$\G$} if all its level graphs are in $\G$. A particularly nice situation occurs when $\G$ is sandwich monotone. In this case, all the edges of an edge-weighted level-$\G$ graph can be eliminated one at a time, from lightest to heaviest, so that all the intermediate graphs are in $\G$, which yields an edge elimination ordering with non-decreasing edge weights. 
Furthermore, we introduce the following strengthening of the concept of sandwich monotonicity:
Given a graph class $\G$, a set $F$ of edges in a graph $G\in \G$ is said to be \emph{$\G$-safe} if $G-F$ is in $\G$.
We call a graph class $\G$ \emph{degree sandwich monotone} if for each graph $ G \in \G$ and each $\G$-safe set $F \subseteq E(G)$, any degree-minimal edge in $F$ is $\G$-safe.
Here an edge $e=uv$ is considered to be \emph{degree-minimal} in $F$ if $u$ has the smallest degree in $G$ among all vertices incident with an edge in $F$ and $v$ has the smallest degree among all vertices that are adjacent to $u$ via an edge contained in $F$.

Given a graph class $\G$, it is natural to ask about the complexity of the recognition of level-$\G$ weighted graphs.
The naive algorithm of checking for each level graph whether it belongs to the respective class can be executed in time $\O(m \cdot p(n,m))$, where $n$ and $m$ are the number of vertices and edges of the input graph and $\O(p(n,m))$ is the complexity of recognizing the unweighted class $\G$. However, for many classes this running time is far from optimal. 
In this work we show that for degree sandwich monotone graph classes there is a necessary and sufficient condition for the existence of a linear-time algorithm for the recognition of level-$\G$ weighted graphs.
This result is achieved by giving a linear-time algorithm which computes a degree-minimal edge elimination scheme of an \emph{arbitrary} weighted graph, i.e., an elimination scheme where each deleted edge is degree-minimal in the set of all edges with minimum weight at the time of deletion.

We apply this general result to weighted analogs of split, threshold, and chain graphs.
In particular, we show that the classes of split, threshold, and chain graphs are degree sandwich monotone, strengthening the previous results about their sandwich monotonicity due to Heggernes and Mancini~\cite{heggernes2009minimal} and Heggernes and Papadopoulos~\cite{heggernes2007single}.
Combining these results with some previous results from the literature we obtain linear-time recognition algorithms of level-split, level-threshold, and level-chain weighted graphs. This is a significant improvement over the naive $\O(m(n+m))$ algorithms.

\paragraph{Related Work.}
Our results related to elimination schemes of weighted graphs can be seen as part of a more general research framework aimed at
generalizing theoretical and algorithmic aspects of graphs to (edge-)weighted graphs, or, equivalently, from binary to real-valued symmetric matrices. For example, Robinsonian similarities are weighted analogues of unit interval graphs~\cite{MR1203643,roberts1969indifference}, Robinsonian dissimilarities
(see~\cite{prea2014optimal}) are weighted analogues of cocomparability graphs~\cite{fortin2017robinsonian}, and similarity-first search is a weighted graph analogue of Lexicographic breadth-first search~\cite{laurent2017similarity}. Other concepts that were generalized to the weighted case include perfect elimination orderings~\cite{laurent2017perfect} and asteroidal triples~\cite{MR3650270}.

All these works, including ours, share a common feature that is often applicable to weighted problems: instead of the exact numerical values of the input, only the structure of these values, that is, the ordinal aspects of the distances or weights, matter. This is a common situation for problems arising in social network analysis (see, e.g.,~\cite{forsyth1946matrix}), combinatorial data analysis (see, e.g.,~\cite{liiv2010seriation}), and phylogenetics (see, e.g.,~\cite{huson1999obtaining,MR1447688}), as well as in greedy algorithms for some combinatorial optimization problems such as Kruskal's or Prim's algorithms for the minimum spanning tree problem, as well as the greedy algorithm for the problem of finding a minimum-weight basis of a matroid~\cite{MR297357}.

Unsurprisingly, concepts similar to those of edge-weighted graphs and their level graphs have appeared in the literature in different contexts and under different names. For example, Berry et al.~\cite{DBLP:journals/soco/BerrySS06} were interested in weighted graphs derived from an experimentally obtained dissimilarity matrix, motivated by questions related to phylogeny reconstruction.
They referred to the family of level graphs as a \emph{threshold family of graphs} and identified a sufficient condition for all the level graphs to be chordal.
The level graphs of a weighted graph can also be seen as a special case of a \emph{temporal graph}, a dynamically changing graph in which each edge can appear and disappear over a certain time period (see, e.g.,~\cite{MR3109184}). In the terminology of Fluschnik et al.~\cite{MR4049937}, the level graphs of a weighted graph form a \emph{1-monotone temporal graph} (see also~\cite{khodaverdian2016steiner}). 
Furthermore, the special case of weighted graphs when all edges have different weights corresponds to an \emph{edge ordered graph}, a concept of interest in extremal graph theory (see, e.g.,~\cite{MR3967297}).

\paragraph{Structure of the Paper.}

In \cref{sec:preliminaries}, we introduce the necessary notation and graph classes considered in this paper. \Cref{sec:weighted} is dedicated to the main structural result of the paper -- a necessary and sufficient condition for linear-time recognition of weighted graph classes.
We use those results later in \cref{sec:safe-characterizations} where we describe the respective recognition procedures for level-split, level-threshold, and level-chain weighted graphs.
\Cref{sec:conclusion} concludes the paper with a short overview and related open questions.

\begin{section}
{Preliminaries}\label{sec:preliminaries}
\end{section}

All graphs considered in this paper are finite, simple, and undirected. Given a graph $G=(V,E)$, we denote by $n = |V|$ and $m = |E|$ the number of vertices and edges in $G$, respectively. A graph is \emph{nontrivial} if it has at least two vertices, and \emph{edgeless} if it contains no edge. 
For a vertex $v$ in a graph $G=(V,E)$, we denote by $N_G(v)$ the \emph{neighborhood} of $v$, i.e., the set $\{u\in V\mid uv\in E\}$, where an edge between $u$ and $v$ in $G$ is denoted by $uv$. 
The \emph{degree} of a vertex $v$ in a graph $G$ is the cardinality of $N_G(v)$ and denoted by $d_G(v)$, and its \emph{closed neighborhood} is the set $N_G[v] = N_G(v)\cup \{v\}$. For a set $X\subseteq V$ we denote by $N_G(X)$ the set of all vertices in $V\setminus X$ that are adjacent to at least one vertex in $X$.
If the graph is clear from the context, we write simply $N(v)$ and $N(X)$ instead of $N_G(v)$ and $N_G(X)$.

A \emph{clique} in a graph $G$ is a set of pairwise adjacent vertices and an
\emph{independent set} in $G$ is a set of pairwise nonadjacent vertices. 
If the neighborhood of a vertex $v$ in $G$ is a clique, then $v$ is said to be a \emph{simplicial vertex}.  
The \emph{complement} of a graph $G$ is the graph $\overline{G}$ with the same vertex set as $G$, where two distinct vertices $u,v\in V$ are adjacent in $\overline{G}$ if and only if they are not adjacent in $G$. 
By $P_n$, $C_n$, and $K_n$ we denote the path, the cycle, and the complete graph of order $n$. In a $P_4$ the edge
  connecting the two vertices with degree two is called the \emph{middle edge} of the
  $P_4$.  A \emph{diamond} is the graph obtained from the $K_4$ by deleting
an edge. In a diamond, we refer to the edge connecting the two degree-three
  vertices as the \emph{middle edge} of the diamond.  A \emph{paw} is a graph
with vertex set $\{a,b,c,d\}$ and edge set
$\{ab,bc,cd,bd\}$. The edges $bc$ and $bd$ are the \emph{side edges} of the paw.  Given two graphs $G$ and $H$, we denote by
$G+H$ their disjoint union, and by $2G$ the disjoint union of two copies of
$G$. In particular, a $2K_2$ is the graph consisting of two disjoint copies of
$K_2$. For a set of graphs $\cal H$ a graph $G$ is called \emph{$\cal H$-free}
if no induced subgraph of $G$ is isomorphic to a graph in $\cal H$.

An \emph{ordering of the vertices} of a graph $G$ is a bijection $\sigma: V(G)\rightarrow \{1,2,\dots,n\}$.
Given two vertices $u$ and $v$ in $G$, we say that $u$ is \emph{to the left} (resp. \emph{to the right}) of $v$ if $\sigma(u)<\sigma(v)$ (resp. $\sigma(u)>\sigma(v)$) and we denote this by $u \prec_{\sigma}v$ (resp.  $u \succ_{\sigma}v$).
Analogously, we define an \emph{ordering of the edges} of $ G $ as a bijection $\tau: E(G) \rightarrow \{1,2,\dots,m\}$.
Given an edge ordering $\tau = (e_1, \ldots, e_m)$ of $G$ we denote by $G^i_\tau$ its spanning subgraph $G - \{e_1, \ldots, e_i\}$.

A graph class is said to be \emph{hereditary}, or \emph{vertex monotone}, if every induced subgraph of
every graph in the class belongs to the class.  
Equivalently, in such a graph class we can delete arbitrary vertices from the graph
and remain in the same class. 
\emph{Edge monotonicity} is defined in the same way: we can
remove arbitrary edges from a graph and remain in the class.  If a class is both edge and vertex monotone, we simply call it \emph{monotone}.  Edge monotonicity is a rather restrictive property and many well-studied graph classes are not edge monotone.

\begin{definition}[{\G-safe edge\,/\,edge set\,/\,edge elimination scheme, grounded graph class}]
Let $\G$ be a graph class and let the graph $G$ be a member of $\G$. An edge $e \in E(G)$ is \emph{\G-safe} if $G - e \in \G$. More generally, a set $F$ of edges of a graph $G\in \G$ is said to be \emph{\G-safe} if $G - F$ is a member of $\G$.
For a graph class $\G$ and a graph $G\in \G$, a \emph{\G-safe edge elimination scheme} of $G$ is defined as an ordering $\tau = (e_1, \ldots, e_m)$ of the edges of $G$ such that for each $i \in \{1,\ldots, m\}$ the spanning subgraph $G^i_\tau$ is in $\G$.
Further, we say that a graph class $\G$ is \emph{grounded} if
for any $G\in \G$ the whole edge set $E(G)$  is $\G$-safe.
\end{definition}
Note that Heggernes and  Papadopoulos~\cite{heggernes2009single} mention the concept of ``edge
  monotonicity'' of a graph class $\G$, meaning that each graph in $\G$ is
  either edgeless or contains an edge $e$ such that $G - e \in \G$.
In the same paper the authors also introduce the concept of \emph{sandwich monotonicity}.

\begin{definition}[{sandwich monotone graph class}]
    A graph class $\G$ is \emph{sandwich monotone if for each graph $G\in \G$ and each
non-empty $\G$-safe set $F \subseteq E(G)$ there is a $\G$-safe edge in
$F$. Equivalently, one can say that a graph class is sandwich monotone if
between any two of its members $G=(V,E)$ and $G' = (V,E \cup F)$ with $E\cap F = \emptyset$ there exists a sequence of graphs $(G = G_0, G_1, \ldots, G_{|F|} = G')$ such that $G_i = G_{i-1} + e$ with $e \in F$ and each graph $G_i$ is a
member of $\G$.}
\end{definition}

 It follows from the second part of the definition that it makes no difference whether we say that we can
delete edges one by one from the larger graph until we get the smaller one
such that all the intermediate graphs are in $\G$, or we consider in a similar way the reverse process of adding edges to the smaller graph until we obtain the larger one. This also leads to the observation already stated
in~\cite{heggernes2009single} that for any sandwich monotone graph class $\G$
the complementary graph class co-$\G := \{\overline{G}\mid G\in \G\}$ is also sandwich monotone.

In the following we present the definitions of the three main graph classes considered in this paper. A \emph{split graph} $G$ is a graph whose vertex set can be partitioned into sets $C$ and $I$ such that $C$ is a clique and $I$ is an independent set in $G$ (see~\cite{MR0505860}). We call $(C,I)$ a \emph{split partition} of $G$.
A graph $ G=(V,E) $ is said to be \emph{threshold} if there exists a labeling $\ell: V \rightarrow \mathbb{N}_0 $ and a threshold value $ t \in \mathbb{N}_0 $ such that a set $ X \subseteq V $ is independent in $G$ if and only if $ \sum_{x \in X} \ell(x) \leq t $ (see~\cite{chvatal1977aggregation}).
The following characterizations of split graphs and threshold graphs are well known.

\begin{theorem}[see, e.g.,~\cite{MR463041}]\label{thm:split}
	A graph $G$ is a split graph if and only if $G$ is $\{2K_2, C_4, C_5\}$-free.
\end{theorem}

\begin{theorem}[see, e.g.,~\cite{mahadev1995threshold}]\label{thm:threshold}
	A graph $G$ is a threshold graph if and only if $G$ is $\{2K_2, P_4,\allowbreak C_4\}$-free.
\end{theorem}

A bipartite graph $G = (V,E)$ is a \emph{chain graph} if its vertex set can be partitioned into two independent sets $X$ and $Y$ such that the vertices in $X$ can be ordered linearly with respect to set inclusion of their neighborhoods (see~\cite{MR604513}). We will refer to such a pair $(X,Y)$ as a \emph{chain bipartition} of $G$.

\begin{theorem}[Hammer, Peled, and Sun~\cite{hammer1990difference}]\label{thm:chain}
A bipartite graph is a chain graph if and only if it is $2K_2$-free.
\end{theorem}

\section{A Sufficient Condition for Linear-Time Recognition of Weighted Graph Classes}\label{sec:weighted}

We proceed with a discussion on sandwich monotonicity in relation to level-$\G$ weighted graphs, followed by an algorithmic approach for their recognition.

\subsection{Level-$\G$ Weighted Graphs: The Definition and Connection with Sandwich Monotonicity}
\label{sec:definition}

In many areas of combinatorics we are given graphs with edge weights of a particular type. For example, in the minimum spanning tree problem we consider only the order of the weights, while the exact weights can be neglected.
This type of weighted graphs plays an important role in the theory of Robinsonian matrices and similarity theory. Here, a symmetric matrix with arbitrary real values can be seen as the adjacency matrix of a weighted graph. Robinsonian matrices $M$ have the special property that for any value $t$ the matrix that results from replacing all values $M_{ij} < t$ with zero and all other values with one is the adjacency matrix of a unit interval graph (as can be easily inferred from~\cite[Theorem 1]{MR1203643}).
Another example is given by the so-called additive matrices, which have the analogous property with respect to chordal graphs (see~\cite{DBLP:journals/soco/BerrySS06}).

A common feature of all these works is that instead of the exact numerical values of the input only the ordinal aspects of the distances or weights matter. This motivates the following definition.

\begin{definition}[{$k$-weighted graph}]
  Given a positive integer $k$, a \emph{$k$-weighted graph} is a pair $(G,\omega)$ where $G$ is a graph and $\omega : E(G) \to \{1,\ldots,k\}$ is a surjective \emph{weight function}.
\end{definition}

Note that a $k$-weighted graph can be thought of as a graph $G$ along with an ordered partition $(E_1,\ldots, E_k)$ of its edge set into $k$ nonempty parts.\footnote{In particular, each $k$-weighted graph gives rise to a symmetric $2$-structure (see, e.g.,~\cite{EHRENFEUCHT1990277}).}
We often denote a $k$-weighted graph $(G,\omega)$ simply by $G$ and call it \emph{weighted}. In applications when a graph is equipped with a weight function of arbitrary real values, sorting the edges by weight yields the required surjective function $\omega:E(G) \to \{1, \ldots, k\}$, where $k$ is the number of distinct weights. Note that in general this requires $\O(m \cdot \log n)$ time if the weights are arbitrary. 

A closely related concept, used in the theory of Robinsonian matrices and similarity theory, is that of \emph{level graphs}, i.e., graphs obtained from an edge-weighted graph by removing all sufficiently light edges.
This suggests a natural way of associating to any class $\G$ of unweighted graphs a corresponding class of edge-weighted graphs, namely by requiring that all level graphs belong to the class.

\begin{definition}[{level graph, level-$\G$ weighted graph}]\label{def:level-graph-sorted}
  Given a weighted graph $(G,\omega)$ with $\omega:E(G) \to \{1, \ldots, k\}$ and an integer $i$ such that $1\le i\le k+1$, the \emph{$i$-th level graph} of $(G,\omega)$ is the graph obtained from $G$ by removing all edges $e$ with $\omega(e) < i$.
  Given a graph class $\G$, we say that a weighted graph $(G,\omega)$ is \emph{level-$\G$} if all level graphs of $G$ are in $\G$.
\end{definition}

Note that the first level graph of a $k$-weighted graph $(G,\omega)$ is $G$ itself and the $(k+1)$-th level graph is the edgeless graph with vertex set $V(G)$.
Recall that a graph class $\G$ is said to be grounded if it is closed under deleting all edges at once.
We were not able to find a natural graph class that is not grounded but contains arbitrarily large edgeless graphs (note, for example, that the class of connected graphs only contains one edgeless graph). For that reason, it seems to be well justified to only consider grounded graph classes in this context.
We show next that under the mild technical condition of groundedness, the sandwich monotonicity of a graph class $\G$ is equivalent to the existence of an edge elimination scheme that is sorted by the weights of the edges.

\begin{definition}[{sorted $\G$-safe edge elimination scheme}]\label{def:sorted-G-safe-edge-elimination-scheme}
  Given a graph class $\G$, a $\G$-safe edge elimination scheme $\tau = (e_1, \ldots, e_m)$ of a weighted graph $G \in \G$ is called \emph{sorted} if for any pair of edges $e_i$ and $e_j$ with $i < j$ it holds that $\omega(e_i) \leq \omega(e_j)$.
\end{definition}

Note that by deleting the edges of a graph $G$ following such a sorted $\G$-safe edge elimination scheme, all level graphs of $G$ are shown to be in $\G$. Thus, the following observation holds.

\begin{observation}\label{obs:safe-level}
  Let $\G$ be a graph class. If a weighted graph $(G, \omega)$ with $G \in \G$ has a sorted $\G$-safe edge elimination scheme, then $(G, \omega)$ is level-$\G$.
\end{observation}

On the other hand, if a grounded graph class $\G$ is sandwich monotone, then any level-$\G$ weighted graph has a sorted $\G$-safe edge elimination scheme.

\begin{proposition}\label{prop:sorted}
  Let $\G$ be a grounded graph class. Then, every level-$\G$ weighted graph has a sorted $\G$-safe edge elimination scheme if and only if $\G$ is sandwich monotone.
\end{proposition}
\begin{proof}
  Assume $\G$ is sandwich monotone. Let $(G,\omega)$ be a level-$\G$ weighted graph with at least one edge and let $F$ be the set of edges of $G$ with minimal weight. Then we know that $G - F$ is still in $\G$. Since $\G$ is sandwich monotone there is a $\G$-safe edge $e$ in $F$. This edge $e$ is the first edge of our ordering  and repeating this argument we find a sorted $\G$-safe edge elimination scheme.

  Now assume that $\G$ is not sandwich monotone. Then there is a graph $G = (V,E)$ in $\G$ and a non-empty $\G$-safe edge set $F \subseteq E$ that does not contain a $\G$-safe edge. We choose the weights $\omega$ of $G$ as follows. All edges in $F$ get weight~$1$ and all other edges get weight~2.
  Since {$F$ is $\G$-safe and} $\G$ is grounded, the weighted graph $(G,\omega)$ is level-$\G$. However, there is no edge with minimal weight which is $\G$-safe. Therefore, there is no sorted $\G$-safe edge elimination scheme of $(G,\omega)$. 
\end{proof}

\subsection{Recognition of Level-$\G$ Weighted Graph Classes}

In the remainder of this section, we use sorted $\G$-safe edge elimination schemes to provide a sufficient condition for the existence of a linear-time recognition algorithm for level-$\G$ weighted graphs (see \cref{thm:general-algo}).
For this we introduce a special case of sandwich-monotonicity that guarantees the existence of a $\G$-safe edge with a special property in a $\G$-safe set. These edges are called degree-minimal.

\begin{definition}[{degree-minimal edge}]\label{def:degree-minimal}
  Let $G$ be a graph and let $F$ be a set of edges of $G$. A \emph{degree-minimal edge in $F$} is an edge $uv\in F$ such that:
  \begin{enumerate}
    \item Vertex $u$ has the smallest degree in $G$ among all vertices incident to an edge in $F$, and
    \item The degree of $v$ in $G$ is the smallest among all neighbors of $u$ that are adjacent to $u$ via an edge in $F$.
  \end{enumerate}
\end{definition}

With this property we can now introduce the degree sandwich monotone graph classes.

\begin{definition}[{degree sandwich monotone graph class}]
  A graph class $\G$ is \emph{degree sandwich monotone} if for each graph $G\in \G$ and each $\G$-safe set $F \subseteq E(G)$ any degree-minimal edge in $F$ is $\G$-safe.
\end{definition}

\begin{figure}[!ht]
  \centering
  \begin{tikzpicture}[xscale=2.5,yscale=1.5]
    \usetikzlibrary{shadows,fadings}
    \tikzset{box/.style={draw,rectangle}}
    \tikzset{roundedbox/.style={draw,rectangle,rounded corners}}
    \scriptsize
    \node[roundedbox] (monotone) at (-1,0) {\mathstrut Monotone};
    \node[box] (degree sandwich monotone) at (-2.9,0.7) {\mathstrut Degree sandwich monotone};
    \node[roundedbox] (hereditary) at (0.6,0.7) {\mathstrut Hereditary};
    \node[roundedbox] (gsm) at (-1,0.7) {\mathstrut Grounded and sandwich monotone};
    \node[roundedbox] (grounded) at (-0.3,1.4) {\mathstrut Grounded};
    \node[roundedbox] (sandwichmonotone) at (-1.7,1.4) {\mathstrut Sandwich monotone};

    \draw[thick,-latex,font=\sffamily] (monotone) to (hereditary);
    \draw[thick,-latex,font=\sffamily] (monotone) to (gsm);
    \draw[thick,-latex,font=\sffamily] (monotone) to (degree sandwich monotone);
    \draw[thick,-latex,font=\sffamily] (gsm) to (sandwichmonotone);
    \draw[thick,-latex,font=\sffamily] (gsm) to (grounded);
    \draw[thick,-latex,font=\sffamily] (degree sandwich monotone) to (sandwichmonotone);
  \end{tikzpicture}
  \caption{Relationships between various monotonicity properties of graph classes. Arrows represent implications, e.g., every monotone graph class is also hereditary.}\label{fig:Hasse}
\end{figure}

Some relationships between the different monotonicity properties of graph classes can be seen in \cref{fig:Hasse}.
Further details will be discussed in our forthcoming paper on relationships between various monotonicity properties of graph classes.

We now present the main result of the section.
This result will be applied several times in~\cref{sec:safe-characterizations}, namely to develop linear-time recognition algorithms for level-split, level-threshold, and level-chain weighted graphs.

\begin{theorem}\label{thm:general-algo}
  Let $\G$ be a grounded degree sandwich monotone graph class. Then there exists a linear-time recognition algorithm for level-$\G$ weighted graphs if and only if the following conditions hold:
  \begin{enumerate}

    \item There exists a linear-time recognition algorithm for graphs in $\G$.

    \item There exists an algorithm which checks in linear time whether a given edge ordering of any graph $G \in \G$ is a $\G$-safe edge elimination scheme of $G$.
  \end{enumerate}
\end{theorem}

In the remainder of this section we  prove this theorem. First we introduce a particular edge elimination scheme of weighted graphs, where each edge is degree-minimal among all edges with minimal weight in the remaining graph. We show that for any degree sandwich monotone graph class $\G$ these elimination schemes characterize the level-$\G$ weighted graphs. At the end of the section we devise a linear-time algorithm that constructs such a scheme for arbitrary weighted graphs. Combining these results proves \cref{thm:general-algo}.

Recall that given an edge ordering $\tau = (e_1, \ldots, e_m)$ of $G$ and an integer $i\in \{1,\ldots, m\}$, we denote by $G^i_\tau$ its spanning subgraph $G - \{e_1, \ldots, e_i\}$, with the special case of $ G^0_{\tau} = G $.

\begin{definition}[{degree-minimal edge elimination scheme}]\label{def:dmees}
  Let $(G,\omega)$ be a weighted graph. A linear ordering $\tau = (e_1, \ldots, e_m)$ of the edges of $G$ is said to be a \emph{degree-minimal edge elimination scheme} of $(G,\omega)$ if for every $i \in \{1,\ldots,m\}$, edge $e_i$ is a degree-minimal edge in the set of all minimum-weight edges in the graph $G^{i-1}_\tau$.
\end{definition}

The next result connects the concepts of degree sandwich monotone graph classes and degree-minimal edge elimination schemes.

\begin{lemma}\label{lemma:degree-elimination}
  Let $\G$ be a grounded degree sandwich monotone graph class and let $(G, \omega)$ be a weighted graph.
  Then $(G, \omega)$ is a level-$\G$ weighted graph if and only if $G \in {\cal G}$ and every degree-minimal edge elimination scheme of $(G, \omega)$ is a sorted $\G$-safe edge elimination scheme.
\end{lemma}
\begin{proof}
  Every weighted graph $(G, \omega)$ has a degree-minimal edge elimination scheme $\tau$. If $\tau$ is a sorted \G-safe edge elimination scheme and $G \in \G$, then $(G, \omega)$ is a level-$\G$ weighted graph, due to \cref{obs:safe-level}.

  Now assume for the other direction that $(G, \omega)$ is a level-$\G$ weighted graph.
  Then $G\in \G$.
  Let $\tau = (e_1, \ldots, e_m)$ be a degree-minimal edge elimination scheme of $(G,\omega)$.
  To prove that $\tau$ is also a sorted $\G$-safe edge elimination scheme of $(G,\omega)$, it is sufficient to show that for all $i\in \{1,\ldots, m\}$, the graph $G^i_\tau$ is in $\G$.
  We prove this by induction on $i$.
  For $i = 1$, we have $G^1_\tau = G-e_1$.
  The edge $e_1$ is degree-minimal in the \G-safe set $F$ of minimum-weight edges in $G$. Since $\G$ is degree sandwich monotone, the edge $e_1$ is $\G$-safe.
  Let $i \in \{2,\ldots, m\}$ and suppose that $G^{i-1}_\tau\in \G$.
  Note that the edge order $(e_i, \ldots, e_m)$ is a degree-minimal edge elimination scheme of
  the weighted graph $(G^{i-1}_\tau, {\omega}{\vert_{E(G^{i-1}_\tau)}})$, where ${\omega}{\vert_{E(G^{i-1}_\tau)}}$ is the restriction of $\omega$ to the edge set of $G^{i-1}_\tau$.
  Thus, $e_{i}$ is a degree-minimal edge in the set of minimum-weight edges in $G^{i-1}_\tau$.
  Again, this set is $\G$-safe and, due to the degree sandwich monotonicity of $\G$, this implies that $e_{i}$ is $\G$-safe.
  Therefore, $G^{i}_\tau$ is in $\G$. 
\end{proof}

The following technical result presents a special data structure that we will
use in our algorithm for computation of a degree-minimal edge elimination scheme
of a weighted graph. A similar data structure was used by Tarjan and Yannakakis~\cite{tarjan1984simple} to implement Maximum Cardinality Search and by Ibarra
in~\cite{ibarra2008fully}. Note that we define the structure as generally as possible.
Therefore, the values $n$ and $k$ have nothing to do with the number of vertices in a graph or with the number of different edges in a weighted
graph.

\begin{lemma}\label{lemma:data-structure}
  Let $k \in \mathbb{N}$ and let $S$ be a given set of $n$ objects with key values in $\{1,\ldots,k\}$. 
  Then, there is a data structure $P$ that fulfills the following conditions:
  \begin{enumerate}
    \item $P$ can be constructed in time $\O(\min\{n\log n, n+k\})$.
    \item $P$ contains an array $A$ whose elements are the objects of $S$ sorted non-decreasingly by their key value.
    \item The following operations can be executed in $\O(1)$ time:
          \begin{itemize}
            \item decrease or increase the key of an object by one,
            \item delete an object with minimal key value or with maximal key value.
          \end{itemize}
  \end{enumerate}
\end{lemma}

\begin{proof}
  We use an array $A$ of size $n$ containing the elements of $S$. At first we sort the elements non-decreasingly with respect to their key values. If $n + k \leq n \log n$ we use counting sort, otherwise we use merge sort. Therefore, we need $\O(\min\{n\log n, n+k\})$ many steps.

  Additionally, we partition the set $S$ into pairwise disjoint non-empty sets $B_j$ where $B_j$ is the set of all objects with key value $j$. Note that each set $B_j$ is a block of consecutive elements of $A$. We represent a block $B_j$ with pointers  $F_j$ and $L_j$ which point to the first and the last element of $B_j$ in $A$, respectively. The tuples $(F_j, L_j)$ are contained in a doubled linked list $D$, which is sorted non-decreasingly by the corresponding key values. Furthermore, every object in $B_j$ has a pointer to the tuple $(F_j, L_j)$.

  \begin{figure}[!ht]
    \centering
    \begin{tikzpicture}[scale=.55]
      \small

      \draw[dashed] (-11.5,-4) -- (-11.5,-13);

      \begin{scope}[yshift=-5.5cm, xscale=-1]
        \foreach \i in {13,...,21} {
            \draw (\i-.5,-.5) rectangle (\i+.5,.5);
            \node at (\i,-.7) {\tiny{\pgfmathparse{34-\i}$\pgfmathprintnumber{\pgfmathresult}$}};
          }

        \node (w) at (14,0) {$v$};
        \node (v) at (16,0) {$w$};
        \draw[->,ultra thick,bend left=30] (w) to (v);
        \draw[->,ultra thick,bend left=30] (v) to (w);

        \fill[gray, fill opacity=0.2,draw=black] (12.5,.5) -- (16.5,.5) -- (16.5,1) arc (0:90:.5) -- (12.5,1.5);
        \draw[very thick,->] (16,-1.5) node[below] {$F_{18}$}-- (16,-1);
        \node at (14,1) {$B_{18}$};

        \fill[gray, fill opacity=0.2,draw=black] (16.5,.5) -- (19.5,.5) -- (19.5,1) arc (0:90:.5) -- (17,1.5) arc (90:180:.5);
        \draw[very thick,->] (19,-1.5) node[below] {$F_{16}$}-- (19,-1);
        \draw[very thick,->] (17,-1.5) node[below] {$L_{16}$}-- (17,-1);
        \node at (18,1) {$B_{16}$};
        \node at (21,1) {$B_{12}$};
        \draw[very thick,->] (20,-1.5) node[below] {$L_{12}$}-- (20,-1);

        \fill[gray, fill opacity=0.2,draw=black] (21.5,1.5) -- (20,1.5) arc (90:180:.5) -- (19.5,.5) -- (21.5,.5);
      \end{scope}

      \begin{scope}[yshift=-11cm, xscale=-1]
        \foreach \i in {13,...,21} {
            \draw (\i-.5,-.5) rectangle (\i+.5,.5);
            \node at (\i,-.7) {\tiny{\pgfmathparse{34-\i}$\pgfmathprintnumber{\pgfmathresult}$}};
          }

        \node at (14,0) {$w$};
        \node at (16,0) {$v$};

        \fill[gray, fill opacity=0.2,draw=black] (12.5,.5) -- (15.5,.5) -- (15.5,1) arc (0:90:.5) -- (12.5,1.5);
        \draw[very thick,->] (15,-1.5) node[below] {$F_{18}$}-- (15,-1);
        \node at (14,1) {$B_{18}$};

        \fill[gray, fill opacity=0.2,draw=black] (16.5,.5) -- (16.5,1) arc (0:180:.5) -- (15.5,.5);
        \draw[very thick,->] (16,-1.5) node[below] {$F_{17}$}-- (16,-1);
        \node at (16,-2.5) {$L_{17}$};
        \node at (16,1) {$B_{17}$};

        \fill[gray, fill opacity=0.2,draw=black] (16.5,.5) -- (19.5,.5) -- (19.5,1) arc (0:90:.5) -- (17,1.5) arc (90:180:.5);
        \draw[very thick,->] (19,-1.5) node[below] {$F_{16}$}-- (19,-1);
        \draw[very thick,->] (17,-1.5) node[below] {$L_{16}$}-- (17,-1);
        \node at (18,1) {$B_{16}$};
        \node at (21,1) {$B_{12}$};
        \draw[very thick,->] (20,-1.5) node[below] {$L_{12}$}-- (20,-1);

        \fill[gray, fill opacity=0.2,draw=black] (21.5,1.5) -- (20,1.5) arc (90:180:.5) -- (19.5,.5) -- (21.5,.5);
      \end{scope}

      \begin{scope}[yshift=-5.5cm,xshift=11cm,xscale=-1]
        \foreach \i in {13,...,21} {
            \draw (\i-.5,-.5) rectangle (\i+.5,.5);
            \node at (\i,-.7) {\tiny{\pgfmathparse{34-\i}$\pgfmathprintnumber{\pgfmathresult}$}};
          }

        \node at (16,0) {$v$};

        \fill[gray, fill opacity=0.2,draw=black] (12.5,.5) -- (15.5,.5) -- (15.5,1) arc (0:90:.5) -- (12.5,1.5);
        \draw[very thick,->] (15,-1.5) node[below] {$F_{18}$}-- (15,-1);
        \node at (14,1) {$B_{18}$};

        \fill[gray, fill opacity=0.2,draw=black] (16.5,.5) -- (16.5,1) arc (0:180:.5) -- (15.5,.5);
        \draw[very thick,->] (16,-1.5) node[below] {$F_{17}$}-- (16,-1);
        \node at (16,-2.5) {$L_{17}$};
        \node at (16,1) {$B_{17}$};

        \fill[gray, fill opacity=0.2,draw=black] (16.5,.5) -- (19.5,.5) -- (19.5,1) arc (0:90:.5) -- (17,1.5) arc (90:180:.5);
        \draw[very thick,->] (19,-1.5) node[below] {$F_{16}$}-- (19,-1);
        \node at (18,1) {$B_{16}$};
        \node at (21,1) {$B_{12}$};
        \draw[very thick,->] (20,-1.5) node[below] {$L_{12}$}-- (20,-1);

        \fill[gray, fill opacity=0.2,draw=black] (21.5,1.5) -- (20,1.5) arc (90:180:.5) -- (19.5,.5) -- (21.5,.5);
      \end{scope}

      \begin{scope}[yshift=-11cm,xshift=11cm,xscale=-1]
        \foreach \i in {13,...,21} {
            \draw (\i-.5,-.5) rectangle (\i+.5,.5);
            \node at (\i,-.7) {\tiny{\pgfmathparse{34-\i}$\pgfmathprintnumber{\pgfmathresult}$}};
          }

        \node at (16,0) {$v$};

        \fill[gray, fill opacity=0.2,draw=black] (12.5,.5) -- (15.5,.5) -- (15.5,1) arc (0:90:.5) -- (12.5,1.5);
        \draw[very thick,->] (15,-1.5) node[below] {$F_{18}$}-- (15,-1);
        \node at (14,1) {$B_{18}$};

        \fill[gray, fill opacity=0.2,draw=black] (16.5,.5) -- (19.5,.5) -- (19.5,1) arc (0:90:.5) -- (16,1.5) arc (90:180:.5)-- (15.5,.5);
        \draw[very thick,->] (19,-1.5) node[below] {$F_{16}$}-- (19,-1);
        \draw[very thick,->] (16,-1.5) node[below] {$L_{16}$}-- (16,-1);
        \node at (18,1) {$B_{16}$};
        \node at (21,1) {$B_{12}$};
        \draw[very thick,->] (20,-1.5) node[below] {$L_{12}$}-- (20,-1);

        \fill[gray, fill opacity=0.2,draw=black] (21.5,1.5) -- (20,1.5) arc (90:180:.5) -- (19.5,.5) -- (21.5,.5);
      \end{scope}
    \end{tikzpicture}
    \caption{Two examples for the data structure update for the decreasing of the key value of object $v$. On the left-hand side, object $v$ with key 18 is first swapped to the beginning of its block. Since there is no object with key 17 (the object left of $v$ has key 16), a new block for key 17 is created. On the right-hand side, the key of $v$ is decreased again. Since $v$ is the first (and only) object in its block, no swap is executed. After decreasing, object $v$ joins the block of objects with key value 16 and block $B_{17}$ is deleted.}\label{fig:degree-structure}
  \end{figure}

  We can construct these data structures in time $\O(n)$. If we want to
  decrease the key of an object $v$ we have to do
  the following procedure (see \cref{fig:degree-structure}). Let
  $j$ be the key of $v$. Then
  $v$ is an element of $B_j$. We swap
  $v$ with the object at position $F_j$ in $A$,
  which means that $v$ is now the first element
  of $B_j$. If the key value of the predecessor of $(F_j,L_j)$ in $D$ is
  equal to $j-1$, then we change the pointer of
  $v$ from $(F_j,L_j)$ to $(F_{j-1},L_{j-1})$ and
  change $L_{j-1}$ to the position of $v$.
  Otherwise, we create a new tuple $(F_{j-1},L_{j-1})$ and insert it
  into $D$ before the tuple $(F_j, L_j)$. The new pointers $F_{j-1}$ and
  $L_{j-1}$ point to the position of $v$.  In
  both cases, we change $F_j$ to the position of the successor of
  $v$ in $A$.  If $F_j$ is now larger than $L_j$,
  then $B_j$ is empty and we remove $(F_j, L_j)$ from the list $D$.  For
  the increasing of the key value of $v$ by one
  we do a similar procedure that swaps $v$ with
  the object at the position $L_j$ and creates or updates the pointers
  of $B_{j+1}$.

  For the deletion of a minimal or maximal element we only have to
  update the first entry of the first element of $D$ or the second entry of the last element of $D$ (and maybe
  delete them from $D$ if the corresponding block $B$ becomes empty).

  It is obvious that we only need a constant number of steps to do all
  of these procedures. Furthermore, it is not hard to see that after
  these procedures the array $A$ is still sorted non-decreasingly by the
  key values and the list $D$ still encodes a valid partition of
  $S$. 
\end{proof}

We now present a linear-time algorithm that computes a degree-minimal edge elimination scheme for an arbitrary weighted graph.

\begin{theorem}\label{thm:degree-algorithm}
  Given a $k$-weighted graph $(G=(V,E), \omega)$, we can compute a degree-minimal edge elimination scheme of $G$ in time $\O(|V|+|E|)$.
\end{theorem}

\begin{proof}
  We prove \cref{thm:degree-algorithm} by describing and analyzing an algorithm with the above properties. We start with the description of the main ideas of the algorithm. For every vertex $v$ and every weight $i$ appearing on an edge incident with $v$ we create a copy of $v$ named $v_i$. Then, we order the vertex copies non-decreasingly with respect to their indices. Here, we use the properties of the data structure to ensure linear running time. The vertex copies with the same index are ordered
  so that the resulting linear order $\sigma$ of all the vertex copies satisfies a certain condition.
  
  For every vertex copy $v_i$ we define the $v_i$-\emph{star} as the edge set
  $\Phi(v_i) = \{vz~|~\omega(vz) = i \text{ and } v_i \prec_\sigma z  _i\}$.
  We create an ordered partition of the edge set based on the order $\sigma$ of the vertex copies by replacing each vertex copy $v_i$ with its respective $v_i$-star $\Phi(v_i)$. Any ordering $\rho = (e_1, \ldots, e_m)$ of the edges of $G$ respecting this partition is sorted with respect to the edge weights and satisfies the following condition: For every edge $e_i$ one of the two incident vertices has minimal degree in $G^{i-1}_\rho$ among all vertices that are incident to an edge with weight $\omega(e_i)$.
  Finally, we reorder the edges within the sets $\Phi(v_i)$ so that condition~2 of \cref{def:degree-minimal} also holds for every edge.

  \paragraph{\bf{Phase 1: Slicing the input graph.}}
  For all $i\in \{1,\ldots, k\}$ we compute the set $V_i$, defined as $V_i := \{v_i\mid v\in V$ is a vertex of $G$ incident to an edge with weight $i\}$ (see \cref{Fig:Ph1} for an example).
  We  refer to $v_i\in V_i$ as the $i$-th \emph{copy} of $v$.
  We denote by $\Xi$ the set $\bigcup_{i = 1}^k V_i$. Note that $|\Xi|\le 2|E|$, since each edge $e = xy\in E$ can generate at most two vertex copies, namely $x_{\omega(e)}$ and $y_{\omega(e)}$.
  For all $i\in \{1,\ldots, k\}$ and all $v_i\in V_i$, we compute the value of $d_i(v)$, where $d_i(v)$ denotes the degree of $v$ in the $i$-th level graph of $(G,\omega)$.

  \begin{figure}[!ht]
    \begin{center}
      \resizebox{\textwidth}{!}{
        \begin{tabular}{|c|c|c|c|c|c|c|c|c|c|c|c|c|c|c|c|c|}
          \hline
          $i$      & \multicolumn{5}{c|}{$1$}                        & \multicolumn{6}{c|}{$2$}                        & \multicolumn{5}{c|}{$3$}                                                                                                                                \\
          \hline
          $G_i$    & \multicolumn{5}{c|}{\begin{tikzpicture}[vertex/.style={inner sep=1.5pt,draw,circle}]
              \begin{scope}
                \node[vertex, label=left: $r$] (1) at (0,0) {};
                \node[vertex, label=above: $s$] (2) at (0.5,0) {};
                \node[vertex, label=above: $t$] (3) at (1,0) {};
                \node[vertex, label=right: $u$] (4) at (1.5,0) {};
                \node[vertex, label=below: $x$] (5) at (0.25,-0.5) {};
                \node[vertex, label=below: $z$] (6) at (1.25,-0.5) {};
                \node[vertex, label=below: $y$] (7) at (0.75,-0.5) {};
                \path[black!50] (1) edge[bend left=75, looseness=1.5] (3);
                \path[black!50] (1) edge[bend left=90, looseness=1.5] (4);
                \path[black!50] (1) edge (5);
                \path[black!50] (2) edge (5);
                \path[black!50] (3) edge (5);
                \path[black!50] (4) edge (5);
                \path[thick] (5) edge (7);
                \path[thick] (3) edge (6);
                \path[thick] (4) edge (6);
                \path[thick] (2) edge[bend left=75, looseness=1.5] (4);
                \path[ultra thick] (1) edge (2);
                \path[ultra thick] (2) edge (3);
                \path[ultra thick] (3) edge (4);
                \path[ultra thick] (1) edge (7);
                \path[ultra thick] (2) edge (7);
                \path[ultra thick] (3) edge (7);
                \path[ultra thick] (4) edge (7);
              \end{scope}
            \end{tikzpicture}} & \multicolumn{6}{c|}{\begin{tikzpicture}[vertex/.style={inner sep=1.5pt,draw,circle}]
              \begin{scope}
                \node[vertex, label=left: $r$] (1) at (0,0) {};
                \node[vertex, label=above: $s$] (2) at (0.5,0) {};
                \node[vertex, label=above: $t$] (3) at (1,0) {};
                \node[vertex, label=right: $u$] (4) at (1.5,0) {};
                \node[vertex, label=below: $x$] (5) at (0.25,-0.5) {};
                \node[vertex, label=below: $z$] (6) at (1.25,-0.5) {};
                \node[vertex, label=below: $y$] (7) at (0.75,-0.5) {};
                \path[thick] (5) edge (7);
                \path[thick] (3) edge (6);
                \path[thick] (4) edge (6);
                \path[thick] (2) edge[bend left=75, looseness=1.5] (4);
                \path[ultra thick] (1) edge (2);
                \path[ultra thick] (2) edge (3);
                \path[ultra thick] (3) edge (4);
                \path[ultra thick] (1) edge (7);
                \path[ultra thick] (2) edge (7);
                \path[ultra thick] (3) edge (7);
                \path[ultra thick] (4) edge (7);
              \end{scope}
            \end{tikzpicture}} & \multicolumn{5}{c|}{\begin{tikzpicture}[vertex/.style={inner sep=1.5pt,draw,circle}]
              \begin{scope}
                \node[vertex, label=left: $r$] (1) at (0,0) {};
                \node[vertex, label=above: $s$] (2) at (0.5,0) {};
                \node[vertex, label=above: $t$] (3) at (1,0) {};
                \node[vertex, label=right: $u$] (4) at (1.5,0) {};
                \node[vertex, label=below: $x$] (5) at (0.25,-0.5) {};
                \node[vertex, label=below: $z$] (6) at (1.25,-0.5) {};
                \node[vertex, label=below: $y$] (7) at (0.75,-0.5) {};
                \path[ultra thick] (1) edge (2);
                \path[ultra thick] (2) edge (3);
                \path[ultra thick] (3) edge (4);
                \path[ultra thick] (1) edge (7);
                \path[ultra thick] (2) edge (7);
                \path[ultra thick] (3) edge (7);
                \path[ultra thick] (4) edge (7);
              \end{scope}
            \end{tikzpicture}}                                                                                                         \\
          \hline
          $V_i$    & $r_1$                                           & $s_1$                                           & $x_1$                                           & $t_1$ & $u_1$ & $x_2$ & $z_2$ & $u_2$ & $t_2$ & $s_2$ & $y_2$ & $r_3$ & $u_3$ & $s_3$ & $t_3$ & $y_3$ \\ \hline
          $d_i(v)$ & $5$                                             & $5$                                             & $5$                                             & $6$   & $6$   & $1$   & $2$   & $4$   & $4$   & $4$   & $5$   & $2$   & $2$   & $3$   & $3$   & $4$   \\
          \hline
        \end{tabular}}
    \end{center}
    \caption{In $G_1$ light gray edges have weight $1$, thin black edges have weight $2$ and thick black edges have weight $3$. Vertices in each $V_i$ are ordered by their non-decreasing degree in $G_i$.}\label{Fig:Ph1}
  \end{figure}

  \paragraph{\bf{Phase 2: Ordering of $\Xi$.}}
  We construct an ordering $\sigma$ of $\Xi$ which respects the fact that degrees of the vertices change during the transition from one level graph to the next, while the edges are eliminated one by one  (see \cref{Fig:Ph2} for an example).
  The ordering $\sigma$ has to fulfill the following properties. First, if $i < j$, then $v_i \prec_\sigma z_j$ for any two vertices $v,z \in V$. Secondly, for any two vertex copies $v_i$ and $z_i$ such that $v_i \prec_\sigma z_i$, the degree of $v$ is smaller than or equal to the degree of $z$ in the graph $G - \bigcup_{x_j \prec_\sigma v_i} \Phi(x_j)$.

  This is achieved by processing $V_i$ one element at a time. Suppose that we have already appended some (possibly none, but not all) elements of $V_i$ to $\sigma$.
  Let $V_i'$ denote the set of remaining elements of $V_i$ and let $Z$ be the set of vertices of $V$ for which there still exist copies in $V_i'$.
  Furthermore, let $F_i$ be the set of edges $vz$ with weight $i$ such that $\{v_i,z_i\}\subseteq V_i'$.
  For all $j>i$ let $F_j$ be the set of edges $vz$ with weight $j$ and let $F =\bigcup_{j\ge i}F_j$.
  We choose a vertex $v$ in $Z$ with smallest degree in the graph $(V,F)$ and append vertex $v_i$ to $\sigma$.

  \begin{figure}[!ht]
    \begin{center}
      \resizebox{0.85\textwidth}{!}{
        \begin{tabular}{|c|x{0.5cm}|x{0.5cm}|x{0.5cm}|x{0.5cm}|x{0.5cm}|x{0.5cm}|x{0.5cm}|x{0.5cm}|x{0.5cm}|x{0.5cm}|x{0.5cm}|x{0.5cm}|}
          \hline
          \makecell{$(V_1,F_1)$} & \multicolumn{5}{c|}{\begin{tikzpicture}[vertex/.style={inner sep=1.5pt,draw,circle}]
              \begin{scope}
                \node[vertex, label=left: $r_1$] (1) at (0,0) {};
                \node[vertex, label=above: $s_1$] (2) at (0.5,0) {};
                \node[vertex, label=above: $t_1$] (3) at (1,0) {};
                \node[vertex, label=right: $u_1$] (4) at (1.5,0) {};
                \node[vertex, label=below: $x_1$] (5) at (0.25,-0.5) {};
                \path[black!50] (1) edge[bend left=75, looseness=1.5] (3);
                \path[black!50] (1) edge[bend left=90, looseness=1.5] (4);
                \path[black!50] (1) edge (5);
                \path[black!50] (2) edge (5);
                \path[black!50] (3) edge (5);
                \path[black!50] (4) edge (5);
              \end{scope}
            \end{tikzpicture}} & \multicolumn{4}{c|}{\begin{tikzpicture}[vertex/.style={inner sep=1.5pt,draw,circle}]
              \begin{scope}
                \node[vertex, label=above: $s_1$] (2) at (0.5,0) {};
                \node[vertex, label=above: $t_1$] (3) at (1,0) {};
                \node[vertex, label=above: $u_1$] (4) at (1.5,0) {};
                \node[vertex, label=below: $x_1$] (5) at (0.25,-0.5) {};
                \path[black!50] (2) edge (5);
                \path[black!50] (3) edge (5);
                \path[black!50] (4) edge (5);
              \end{scope}
            \end{tikzpicture}} & \multicolumn{3}{c|}{\begin{tikzpicture}[vertex/.style={inner sep=1.5pt,draw,circle}]
              \begin{scope}
                \node[vertex, label=above: $s_1$] (2) at (0.5,0) {};
                \node[vertex, label=above: $t_1$] (3) at (1,0) {};
                \node[vertex, label=above: $u_1$] (4) at (1.5,0) {};
                \node[label=below: $\,$] (5) at (0.25,-0.5) {};
              \end{scope}
            \end{tikzpicture}}                                                                                                \\
          \hhline{*{13}{-}}
          $V_1'$                 & {\cellcolor{black!15}}$r_1$                    & $s_1$                                           & $x_1$                                           & $t_1$ & $u_1$ & $s_1$ & {\cellcolor{black!15}}$x_1$ & $t_1$ & $u_1$ & $s_1$ & $t_1$ & $u_1$ \\
          \hhline{*{13}{-}}
          $d_{(V,F)}(v_i)$       & {\cellcolor{black!15}}$5$                      & $5$                                             & $5$                                             & $6$   & $6$   & $5$   & {\cellcolor{black!15}}$4$   & $4$   & $4$   & $4$   & $3$   & $3$   \\
          \hline
          $\sigma$               & \multicolumn{5}{c|}{$r_1$}                      & \multicolumn{4}{c|}{$x_1$}                      & \multicolumn{3}{c|}{$t_1,u_1,s_1$}                                                                                                             \\
          \hline
        \end{tabular}}
    \end{center}
    \caption{Processing of the first slice of $\Xi$ in Phase $2$. After all slices have been processed, we get $\sigma=(r_1,x_1,t_1,u_1,s_1,x_2,z_2,u_2,t_2,s_2,y_2,r_3,u_3,y_3,t_3,s_3)$. Note that at the beginning of the process we can choose any vertex $v_i$ with $d_{(V,F)}(v_i)=5$, and in each case a different ordering $\sigma$ is obtained.}\label{Fig:Ph2}
  \end{figure}

  \paragraph{\bf{Phase 3: Ordering the edges.}} The linear order $\sigma$ induces an ordered partition of the edges of $G$ by replacing each vertex copy $v_i$ with its respective $v_i$-star $\Phi(v_i)$ in $\sigma$. Any ordering $\rho = (e_1, \ldots, e_m)$ of the edges of $G$ respecting this partition is sorted with respect to the edge weights and satisfies the following condition: For every edge $e_i$, one of the two incident vertices has minimal degree in $G^{i-1}_\rho$ among all vertices that are incident to an edge with weight $\omega(e_i)$. However, such an order $\rho$ does not necessarily satisfy condition~2 of \cref{def:degree-minimal} (see \cref{Fig:Ph3} for an example). This phase of the algorithm computes the final edge order $\tau$ of the edges by sorting the elements of the $v_i$-stars.
  For every $i \in \{1,\ldots, k\}$ and every copy $v_i$ we order the edges $vz$ of the $v_i$-star non-decreasingly with respect to the degree of $z$ in the graph $(V,F)$, where $F$ is the union of all $z_j$-stars where $z_j = v_i$ or $v_i\prec_{\sigma} z_j$.

  \begin{figure}[!ht]
    \begin{center}
      \resizebox{\textwidth}{!}{
        \begin{tabular}{|c|c|c|c|c|c|c|c|c|c|c|c|c|c|c|c|c|c|}
          \hline
          \makecell{$(V,F)$}           & \multicolumn{3}{c|}{\begin{tikzpicture}[vertex/.style={inner sep=1.5pt,draw,circle},scale=0.85]
              \begin{scope}
                \node[vertex, label=above: $r$] (1) at (0,0) {};
                \node[vertex, label=above: $s$] (2) at (0.5,0) {};
                \node[vertex, label=above: $t$] (3) at (1,0) {};
                \node[vertex, label=above: $u$] (4) at (1.5,0) {};
                \node[vertex, label=below: $x$] (5) at (0.25,-0.5) {};
                \node[vertex, label=below: $z$] (6) at (1.25,-0.5) {};
                \node[vertex, label=below: $y$] (7) at (0.75,-0.5) {};
                \path[black!50] (1) edge[bend left=75, looseness=1.5] (3);
                \path[black!50] (1) edge[bend left=90, looseness=1.5] (4);
                \path[black!50] (1) edge (5);
                \path[black!50] (2) edge (5);
                \path[black!50] (3) edge (5);
                \path[black!50] (4) edge (5);
                \path[thick] (5) edge (7);
                \path[thick] (3) edge (6);
                \path[thick] (4) edge (6);
                \path[thick] (2) edge[bend left=75, looseness=1.5] (4);
                \path[ultra thick] (1) edge (2);
                \path[ultra thick] (2) edge (3);
                \path[ultra thick] (3) edge (4);
                \path[ultra thick] (1) edge (7);
                \path[ultra thick] (2) edge (7);
                \path[ultra thick] (3) edge (7);
                \path[ultra thick] (4) edge (7);
              \end{scope}
            \end{tikzpicture}} & \multicolumn{3}{c|}
          {\begin{tikzpicture}[vertex/.style={inner sep=1.5pt,draw,circle},scale=0.85]
              \begin{scope}
                \node[vertex, label=above: $r$] (1) at (0,0) {};
                \node[vertex, label=above: $s$] (2) at (0.5,0) {};
                \node[vertex, label=above: $t$] (3) at (1,0) {};
                \node[vertex, label=above: $u$] (4) at (1.5,0) {};
                \node[vertex, label=below: $x$] (5) at (0.25,-0.5) {};
                \node[vertex, label=below: $z$] (6) at (1.25,-0.5) {};
                \node[vertex, label=below: $y$] (7) at (0.75,-0.5) {};
                \path[black!50] (2) edge (5);
                \path[black!50] (3) edge (5);
                \path[black!50] (4) edge (5);
                \path[thick] (5) edge (7);
                \path[thick] (3) edge (6);
                \path[thick] (4) edge (6);
                \path[thick] (2) edge[bend left=75, looseness=1.5] (4);
                \path[ultra thick] (1) edge (2);
                \path[ultra thick] (2) edge (3);
                \path[ultra thick] (3) edge (4);
                \path[ultra thick] (1) edge (7);
                \path[ultra thick] (2) edge (7);
                \path[ultra thick] (3) edge (7);
                \path[ultra thick] (4) edge (7);
              \end{scope}
            \end{tikzpicture}} & \multicolumn{8}{c|}{$\cdots$}                   & \multicolumn{2}{c|}{\begin{tikzpicture}[vertex/.style={inner sep=1.5pt,draw,circle},scale=0.85]
              \begin{scope}
                \node[vertex, label=above: $t$] (2) at (0.5,0) {};
                \node[vertex, label=above: $u$] (3) at (1,0) {};
                \node[vertex, label=below: $x$] (7) at (0.75,-0.5) {};
                \path[ultra thick] (2) edge (3);
                \path[ultra thick] (2) edge (7);
                \path[ultra thick] (3) edge (7);
              \end{scope}
            \end{tikzpicture}} & \begin{tikzpicture}[vertex/.style={inner sep=1.5pt,draw,circle},scale=0.85]
            \begin{scope}
              \node[vertex, label=above: $t$] (2) at (0.5,0) {};
              \node[vertex, label=above: $u$] (3) at (1,0) {};
              \node[label=below: $\,$] (7) at (0.75,-0.5) {};
              \path[ultra thick] (2) edge (3);
            \end{scope}
          \end{tikzpicture}                                                                                                                                                                                                                                                                             \\
          \hline
          $\Phi(v_i)$                  & \multicolumn{3}{c|}{$\Phi(r_1)$}                & \multicolumn{3}{c|}{$\Phi(x_1)$}                & $\Phi(x_2)$                & \multicolumn{2}{c|}{$\Phi(z_2)$} & $\Phi(u_2)$ & \multicolumn{2}{c|}{$\Phi(r_3)$} & \multicolumn{2}{c|}{$\Phi(u_3)$} & \multicolumn{2}{c|}{$\Phi(y_3)$} & $\Phi(t_3)$                                                                                                     \\
          \hline
          $\rho$                       & $rx$                                            & $rt$                                            & $ru$                       & $xs$                             & $xt$        & $xu$                             & $xy$                             & $zu$                             & $zt$        & $us$ & $ry$                       & $rs$                       & $uy$ & $ut$ & $yt$ & $ys$ & $ts$ \\
          \hline
          $d_{(V,F)}(w)$               & $5$                                             & $6$                                             & $6$                        & $5$                              & $5$         & $5$                              & $5$                              & $4$                              & $4$         & $4$  & {\cellcolor{black!15}}$4$ & {\cellcolor{black!15}}$3$ & $3$  & $3$  & $2$  & $2$  & $1$  \\
          \hline
          $\tau$                       & $rx$                                            & $rt$                                            & $ru$                       & $xs$                             & $xt$        & $xu$                             & $xy$                             & $zu$                             & $zt$        & $us$ & $rs$                       & $ry$                       & $uy$ & $ut$ & $yt$ & $ys$ & $ts$ \\
          \hline
        \end{tabular}}
    \end{center}
    \caption{In Phase $3$ the vertex ordering from Phase~2 is used to compute the desired edge elimination ordering. 
    Note that even though $rs \prec_\rho ry$, the edge $ry$ appears before $rs$ in $\tau$. In the last two graphs $(V,F)$, vertices of degree $0$ are not depicted.}\label{Fig:Ph3}
  \end{figure}

  \paragraph{\bf{Implementation details.}} We  now describe how we can achieve a linear running time. First we sort the edges in $E$ non-decreasingly according to their weights, which can be done in time $\O(|E|)$ using counting sort. To create $\Xi$, we store for every vertex the copy created last. We traverse the edges according to their order. If for one of the vertices incident to the current edge $e$ there is no copy for the weight of $e$, we create it. All edges are assigned a pointer to their corresponding vertex copies. This process can be done in linear time.

  The values $d_i(v)$ can be computed in linear time by traversing the edges according to their order and updating the degrees. We assign the value $d_i(v)$ to $v_i$. Since the degrees lie between 0 and $|V|-1$, we can order all vertex copies in linear time with counting sort with regard to the values $d_i$ and then place them in their corresponding sets $V_i$ in the order of their degrees.

  In Phase~2 we use a data structure of \cref{lemma:data-structure} for every slice of $\Xi$ separately, where the copies correspond to the objects and the values $d_i$ are used as key values. As described above, we can sort all the slices in total time $\O(|V| + |E|)$. Since the slices are disjoint and have an overall size in $\O(|E|)$, the remaining operations for the construction of the data structures can also be done in linear time. Furthermore, the vertex copy $v_i$ with minimal degree can be found in constant time and updating the degrees of the other copies when the $v_i$-star is deleted only costs linear time overall.

  In Phase~3 we compute for every edge $e= vz$ with
  $v_{\omega(e)} \prec_\sigma z_{\omega(e)}$
  the degree of $z$ in the graph obtained by deleting all $x_i$-stars with $x_i \prec_\sigma v_{\omega(e)}$. As in Phase~1, this can be done in linear time by traversing the $x_i$-stars with respect to $\sigma$. Afterwards, we sort all edges with respect to the computed degrees in linear time with counting sort and reinsert them into their $v_i$-stars according to the computed degree ordering, leading to the desired edge ordering $\tau$. 
\end{proof}

Now we are able to prove \cref{thm:general-algo}.

\begin{proof}[Proof of \cref{thm:general-algo}]
  Assume that conditions 1 and 2 from the theorem hold and let $(G, \omega)$ be a weighted graph to be tested for the property of being level-$\G$. We first check with the given linear-time recognition algorithm of $\G$ whether $G$ is an element of $\G$. If this is not the case, then $(G, \omega)$ is not level-\G. Otherwise, we compute a degree-minimal edge elimination scheme $\sigma$ using the algorithm given by \cref{thm:degree-algorithm}. This needs time linear in the size of $G$.
  If $\sigma$ is a sorted \G-safe edge elimination scheme, then $(G, \omega)$ is level-$\G$ by \cref{obs:safe-level}. Otherwise, $(G, \omega)$ is not level-$\G$ due to \cref{lemma:degree-elimination}.

  For the other direction suppose that we are given a recognition algorithm $\mathcal A$ for the level-$\G$ weighted graphs.
  We begin by showing that there exists a linear-time recognition algorithm for $\G$.
  To this end, let $G$ be an arbitrary graph.
  The function $\omega$ is defined by assigning weight $1$ to all edges of $G$. Now we apply $\mathcal A$ to $(G,\omega)$.
  If $\mathcal A$ decides that $(G,\omega)$ is a level-$\G$ weighted graph, then $G\in \G$.
  Otherwise one of the two level graphs ($G$ or $G-E(G)$) is not a member of $\G$.
  As $\G$ is grounded, this implies that $G\notin \G$.
  This procedure defines a linear recognition algorithm for the given graph class $\G$.
  It remains to show that there exists an algorithm which checks in linear time whether a given edge ordering is a $\G$-safe edge elimination scheme.
  Now let $G$ be a member of $\G$ and let $\sigma=(e_1,\dots,e_m)$ be an arbitrary ordering of the edges of $G$.
  We define the weight function $\omega$ as
  $\omega(e_i) = i$ for all $i\in \{1,\dots, m\}$.
  We apply $\mathcal A$ to $(G,\omega)$.
  The algorithm $\mathcal A$ returns true if and only if every level graph of $(G,\omega)$ is a member of $\G$.
  As $\G$ is grounded, this implies that $\sigma$ is a $\G$-safe edge elimination ordering of $G$. 
\end{proof}

\section{Recognizing Level-Split, Level-Threshold, and Level-Chain Weighted Graphs}\label{sec:safe-characterizations}

The main result of this section is to establish the existence of linear-time recognition algorithms for the classes of level-$\G$ weighted graphs, when $\G$ is one of the following: the class of split graphs, the class of threshold graphs, or the class of chain graphs.
For this we apply \cref{thm:general-algo} to the graph class $\G$.
Two challenges arise: showing that $\G$ is degree sandwich monotone and finding an algorithm which can decide whether a given edge elimination scheme is $\G$-safe.

Finding an algorithm that can decide whether a given edge elimination scheme is $\G$-safe can be done by using a \emph{dynamic recognition algorithm} of a graph class $\G$, which is defined as follows. For a given graph
  $G \in \G$ the algorithm may first execute a preprocessing phase (for example, to compute some special data structure). 
  We call the time that the
  algorithm needs for this task the \emph{preprocessing time}.
  Afterwards, the algorithm gets a sequence of edge deleting operations, which is supposed to
  be applied on $G$.  As long as the graph stays in the graph class \G, the algorithm
  must be able to delete the respective edge $e$ from the remaining graph
  $G'$. We call the time that the algorithm needs to check whether the graph
  $G' - e$ is still in $\G$ and if so, for deleting $e$ from $G'$, the
  \emph{deletion time}.
\begin{observation}\label{obs:dynamic-edge-schemes}
    Let $\G$ be a graph class.
    If there is a dynamic recognition algorithm for $\G$ with linear preprocessing time and constant deletion time, then there exists a linear-time algorithm which decides whether a given edge ordering of a graph $G\in\G$ is a $\G$-safe edge elimination scheme.
\end{observation}

To show the degree sandwich monotonicity for each of the three classes, our approach is to first characterize the set of safe edges and then show that every degree-minimal edge in a safe set is safe.
In both cases we make use of the forbidden induced subgraph characterization of the corresponding graph class.
This gives a unifying way to obtain alternative short proofs of the sandwich monotonicity of the considered graph classes, which was established already in~\cite{heggernes2009minimal,heggernes2009single}.

We first observe a simple but useful general property of degree-minimal edges.

\begin{lemma}\label{lem:degree-minimal-neighbors}
  Let $G$ be a graph, $F$ be a set of edges of $G$, and $xy$ a degree-minimal edge in $F$.
  Then, for every edge $xz\in F$, we have $d_{G}(z)\ge d_{G}(y)$.
\end{lemma}

\begin{proof}
  If $d_{G}(x)\le d_{G}(y)$, then vertex $x$ has the smallest degree in $G$ among all vertices incident to an edge in $F$, and hence the inequality holds by the second condition of the definition of a degree-minimal edge. If $d_{G}(x) > d_{G}(y)$, then vertex $y$ has the smallest degree in $G$ among all vertices incident to an edge in $F$, and hence $d_{G}(z)\ge d_{G}(y)$. 
\end{proof}

We now focus our attention to individual graph classes.

\subsection{Split Graphs}\label{sec:split}

First we characterize split-safe edges. Recall that the middle edge of a $P_4$ is the edge connecting the two vertices with degree two, while the middle edge of a diamond is the edge connecting the two vertices with degree three.

\begin{lemma}\label{lemma:split-safe1}
  Let $G= (V,E)$ be a split graph. An edge $e \in E$ is split-safe if and only if it is neither the middle edge of an induced $P_4$ nor the middle edge of an induced diamond in $G$.
\end{lemma}

\begin{proof}
  Recall by \cref{thm:split} that a graph is a split graph if and only if it is $\{2K_2,C_4,C_5\}$-free.
  Assume $e$ is the middle edge of an induced $P_4$. If we delete $e$, then we obtain an induced $2K_2$ in the graph, which implies that $G-e$ is not split. If $e$ is the middle edge of an induced diamond, then its deletion creates an induced $C_4$ and, thus, $G - e$ is not split.

  On the other hand, an edge $e$ is not split-safe if and only if its deletion creates an induced copy of either $C_4$, $C_5$, or $2K_2$. If we obtain a $C_4$, then $e$ was the middle edge of an induced diamond. If we obtain a $C_5$, then $G$ already contained an induced $C_4$, which is not possible in a split graph. If we obtain a $2K_2$, then $e$ was the middle edge of an induced $P_4$. 
\end{proof}

\begin{lemma}
    Let $G = (V,E)$ be a split graph and $F \subseteq E$ be a set such that $G - F$ is also a split graph. Let $e \in F$ be an edge incident to a vertex of minimum degree among all vertices that are incident to an edge of $F$. Then $G - e$ is also a split graph.
\end{lemma}

\begin{proof}
  Fix a split partition $(C,I)$ of $G$. Suppose $e = xy$ is not split-safe. By \cref{lemma:split-safe1}, $e$ is either the middle edge of an induced $P_4$ or the middle edge of an induced diamond in $G$.

  Suppose first that $e$ is the middle edge of an induced $P_4$, say
  $(u,x,y,v)$. Then $x,y\in C$ and $u,v\in I$. Since the graph $G' = G-F$ is a
  split graph and hence $2K_2$-free, at least one of the edges $ux$ and $vy$
  belongs to $F$. Without loss of generality, we may assume that $ux\in F$. 
  Since $u\in I$ and $x,y\in C$, we have $N_G(u)\subseteq C\setminus \{y\}\subseteq N_G[x]\cap N_G(y)$. 
  Hence, since both $x$ and $y$ have a neighbor that is not in $N_G[u]$, we obtain that $d_G(x) > d_G(u)$ and $d_G(y) > d_G(u)$, a contradiction to the choice of $e$.

  Thus, we may assume that $e$ is the middle edge of an induced diamond with vertex
  set $\{u,v,x,y\}$.  Since $C$ is a clique in $G$, at least one of the
  vertices $u$ and $v$ belongs to $I$. We may assume without loss of generality
  that $u\in I$. This implies that every neighbor of $u$ is in $C$; in
  particular, $x,y\in C$. As the degree of every vertex in $I$ is strictly smaller than $\min(d_{G}(x),d_{G}(y))$, none of the edges between a vertex in $I$ and a vertex in $C$ is an element of $F$.
  In particular, none of the edges $ux$ and $uy$ is
  in $F$. Since the graph $G' = G-F$ is a split graph and hence $C_4$-free, at
  least one of the edges $vx$ and $vy$ belongs to $F$. By symmetry, we may
  assume that $vx\in F$.
  By the arguments above, vertex $v$ cannot belong to $I$
    since that would imply that $d_G(v) < \min(d_{G}(x),d_{G}(y))$, contradicting the choice of $e$. Thus, $v\in C$.

\begin{figure}[!ht]
  \centering
  \begin{tikzpicture}[vertex/.style={inner sep=1.5pt,draw,circle}, scale=0.7]
    \begin{scope}
      \tikzset{decoration={snake,amplitude=.4mm,segment length=2mm,%
          post length=0mm,pre length=0mm}}%

      \node[vertex, label=left: $u$] (1) at (0,0) {};%
      \node[vertex, label=above: $x$] (2) at (2,1) {};%
      \node[vertex, label=below: $y$] (3) at (2,-1) {};%
      \node[vertex, label=north east: $v$] (4) at (4,0) {};%
      \node[vertex, label=below: $t$] (5) at (5,-1) {};%

      \path (1) edge[thick] (2);%
      \path (1) edge[thick] (3);%
      \path (3) edge[thick] (4);%
      \path (4) edge[thick] (5);%
      \path (2) edge[thick,dashed,bend left=60, looseness=1.5] (5);%

      \path (2) edge[thick,decorate] (4);%
      \path (2) edge[thick,decorate] (3);%
    \end{scope}
  \end{tikzpicture}
  \caption{Picture visualizing the proof of \cref{thm:split-degreeminimal}.
  In the case of $d_{G'}(v) \geq d_{G'}(x)$, the dashed edge $xt$ is not
    there, the wavy edges are contained in $F$, and all undecorated edges apart from
    $yv$ cannot be contained in $F$.\\In the
    case of $d_{G'}(v) < d_{G'}(x)$, the dashed edge $xt$ might be there, the
    wavy edges are contained in $F$, and the straight edges cannot be in $F$.}\label{fig:proof}
\end{figure}

Assume first that
    $d_{G}(v)\ge d_{G}(x)$.
    Since $u$ is adjacent to $x$ but not to $v$,
  there exists a vertex $t\in V(G)\setminus\{u,v,x,y\}$ such that $t$ is
  adjacent to $v$ but not to $x$. Since $v\in C$, we have $t\in I$.
  Furthermore, we must have $tv\notin F$, and thus the subgraph of $G'$ induced by
  $\{t,u,v,x\}$ is isomorphic to $2K_2$; a contradiction to the fact that
  $G' = G-F$ is a split graph (see \cref{fig:proof}).

  Now assume that $d_{G}(v)< d_{G}(x)$. The choice of $e$ implies that $d_G(y) \leq d_G(v)$. 
  Similarly as in the above case, there exists a vertex $t\in V(G)\setminus\{u,v,x,y\}$ such that $t$ is adjacent to $v$ but not to $y$.
    Since $v\in C$, we have $t\in I$ and $tv\notin F$. Now observe that, depending on whether $xt\in E(G)$ or not, the set $\{x,u,y,v,t\}$ either forms an induced $P_5$, or an induced $C_5$ in $G$. In either case we obtain a contradiction to the fact that
  $G' = G-F$ is a split graph (see \cref{fig:proof}). 
\end{proof}

As the degree-minimal edges are a subset of the edges considered in the above lemma, degree-minimal edges in $F$ are split-safe. This implies the following.

\begin{theorem}\label{thm:split-degreeminimal}
  The class of split graphs is degree sandwich monotone.
\end{theorem}

\begin{theorem}[Hammer and Simeone~\cite{hammer1981splittance}]\label{thm:split-recognition}
  There exists a linear-time recognition algorithm for split graphs.
\end{theorem}

\begin{theorem}[Ibarra~\cite{ibarra2008fully}]\label{thm:dynamic-split-recognition}
  There exists a dynamic recognition algorithm for split graphs with linear preprocessing time and constant deletion time.
\end{theorem}

Combining \cref{thm:split-degreeminimal,thm:split-recognition,thm:dynamic-split-recognition} with \cref{obs:dynamic-edge-schemes,thm:general-algo}  we
obtain a linear-time algorithm that decides whether a given weighted graph $(G,\omega)$ is a level-split graph.

\begin{theorem}\label{thm:level-G-recognition-split}
  Let $\G$ be the class of split graphs. 
  Then there exists a linear-time recognition algorithm for level-$\G$ weighted graphs.
\end{theorem}

\subsection{Threshold Graphs}\label{sec:threshold}

We begin by characterizing threshold-safe edges. Recall that a side edge of a paw is an edge connecting the vertex of degree three with a vertex of degree two.

\begin{lemma}\label{lemma:threshold}
  Let $G= (V,E)$ be a threshold graph. An edge $e \in E$ is threshold-safe if and only if it is neither the middle edge of an induced diamond nor a side edge of  an induced paw in $G$.
\end{lemma}

\begin{proof}
  Recall by \cref{thm:threshold} that a graph is a threshold graph if and only
  if it is $\{2K_2,P_4,C_4\}$-free.  Deleting the middle edge of an induced
  diamond we obtain an induced $C_4$, while
  deleting a side edge in a paw we obtain an
  induced subgraph isomorphic to $P_4$. Thus, such edges cannot be
  threshold-safe.

  Assume an edge $e\in E$ is not threshold-safe. Then the removal of $e$
  creates an induced subgraph isomorphic to $2K_2$, $P_4$, or $C_4$.  First, observe that a $2K_2$ may only be obtained by removing the middle edge of a
  $P_4$, which cannot appear in a threshold graph.  Next, notice that the
  $C_4$ can only be obtained by removing the middle edge of a diamond.  Finally,
  as a threshold graph cannot contain an induced $C_4$, the only way to obtain
  an induced subgraph isomorphic to $P_4$ is by deleting a side edge in a
  paw. 
\end{proof}

Before showing that threshold graphs are degree sandwich monotone, we establish the following lemma.

\begin{lemma}\label{lem:threshold-degree}
  Let $G$ be a threshold graph and
  let $u,v\in V(G)$ be two vertices in $G$ such that
  $N_G(v)\nsubseteq N_G[u]$. Then $d_G(v)>d_G(u)$.
\end{lemma}

\begin{proof}
  It suffices to show that every vertex adjacent to $u$ is either equal or
  adjacent to $v$. Suppose this is not the case, that is, there exists a
  vertex $w\neq v$ that is adjacent to $u$ but not adjacent to $v$. By
  assumption, there exists a vertex $z\neq u$ that is adjacent to $v$ but not
  to $u$.  Then $vz$ and $uw$ are edges of $G$ and $vw$ and $uz$ are non-edges
  of $G$. Since the four vertices $u,v,w,z$ are pairwise distinct, the
  subgraph of $G$ induced by $\{u,v,w,z\}$ is isomorphic to either $P_4$,
  $2K_2$, or $C_4$. By \cref{thm:threshold}, this contradicts the fact
  that $G$ is threshold. 
\end{proof}

\begin{theorem}\label{thm:threshold-degreeminimal}
  The class of threshold graphs is degree sandwich monotone.
\end{theorem}

\begin{proof}
  {Let $G = (V,E)$ and $G' = (V,E\cup F)$ be threshold graphs, where $E\cap F = \emptyset$.
    We need to show that every degree-minimal edge in $F$ is threshold-safe.}
  Suppose for a contradiction that some degree-minimal edge in $F$, say
  $e = xy$, is not threshold-safe.  By \cref{lemma:threshold}, $e$ is
  either the middle edge of an induced diamond or a
  side edge in an
  induced paw in $G'$.

  Suppose first that $e$ is the middle edge of a diamond with vertex set
  $\{u,v,x,y\}$. Since $G = G'-F$ is threshold and hence $C_4$-free, $e$
  cannot be the only edge of the diamond that belongs to $F$. By symmetry, we
  may assume that $ux\in F$. Since $v\in N_{G'}(x)\setminus N_{G'}[u]$,
  \cref{lem:threshold-degree} implies that
  $d_{G'}(u)<d_{G'}(x)$. Similarly, $d_{G'}(u)<d_{G'}(y)$. This contradicts
  the degree-minimality of $e=xy$.

  Suppose now that $e=xy$ is a
  side edge in a
  paw with vertex set $\{u,v,x,y\}$, where $v$ is the vertex of degree one in
  the paw and $x$ the degree two vertex of $e$. Since
  $v\in N_{G'}(y)\setminus N_{G'}[x]$ and $u\in N_{G'}(x)\setminus N_{G'}[v]$,
  \cref{lem:threshold-degree} implies that
  $d_{G'}(v)<d_{G'}(x)<d_{G'}(y)$. Similarly, $d_{G'}(u)<d_{G'}(y)$. Since
  $G = G'-F$ is threshold and hence $\{P_4,2K_2\}$-free, at least one of the
  edges $ux$ and $vy$ belongs to $F$. If $ux\in F$, then inequalities
  $d_{G'}(x)<d_{G'}(y)$ and $d_{G'}(u)<d_{G'}(y)$ contradict the
  degree-minimality of $e = xy$. Therefore, $vy\in F$. But now, the
  inequalities $d_{G'}(v)<d_{G'}(x)<d_{G'}(y)$ contradict the
  degree-minimality of $e$. 
\end{proof}

Note that in contrast to split graphs, we actually need that the edge is degree-minimal, i.e., it is not sufficient that one of the incident vertices has minimal degree among all vertices incident to edges of $F$. An example is given in \cref{fig:counterexample}.

The proof of \cref{thm:threshold-degreeminimal} implies a property that seems to be even stronger than degree sandwich monotonicity. If a forbidden induced subgraph is constructed when deleting a degree-minimal edge in $F$, then this subgraph is also contained in $G - F$.

\begin{corollary}\label{cor:theshold-certificate}
    Let $G$ be a threshold graph and $F \subseteq E(G)$. If the degree-minimal edge $e$ in $F$ is not threshold-safe, then every induced $2K_2$, $P_4$, and $C_4$ in $G - e$ is also contained in $G - F$.
\end{corollary}%
\begin{figure}
    \centering
    \begin{tikzpicture}[vertex/.style={inner sep=1.5pt,draw,circle}, scale=1.5]
    \tikzset{decoration={snake,amplitude=.4mm,segment length=2mm,%
          post length=0mm,pre length=0mm}}%
        \node[vertex, label=left: $u$] (1) at (0,0) {};
        \node[vertex, label=right: $v$] (2) at (1,0) {};
        \node[vertex, label=left: $w$] (3) at (0,1) {};
        \node[vertex, label=right: $x$] (4) at (1,1) {};

        \draw[thick] (1) -- (3) -- (4);
        \draw[thick, decorate] (2) -- (3);
        \draw[thick, decorate] (2) -- (4);
    \end{tikzpicture}
    \caption{The graph consisting of the straight and wavy edges is a threshold graph. Let $F$ be the set of wavy edges. Then $G - F$ is also a threshold graph. Vertex $v$ has minimal degree among all vertices incident to edges of $F$. Nevertheless, edge $vw$ is not threshold-safe.}
    \label{fig:counterexample}
\end{figure}%

\begin{theorem}[Chv\'atal and Hammer~\cite{chvatal1977aggregation}]\label{thm:threshold-recognition}
  There exists a linear-time recognition algorithm for \mbox{threshold} graphs.
\end{theorem}

Shamir and Sharan~\cite{shamir2004fully} presented a dynamic recognition algorithm for threshold graphs with linear preprocessing time and constant deletion time. Beisegel et al.~\cite{MR4624523} improved this result to a certifying algorithm.

\begin{theorem}[{Beisegel et al.~\cite{MR4624523}}]\label{thm:dynamic-threshold-recognition}
  There exists a dynamic recognition algorithm for threshold graphs with linear preprocessing time and constant deletion time. If the graph $G'$ that results from the deletion of an edge is not a threshold graph, then the algorithm returns an induced $2K_2$, $P_4$ or $C_4$ of $G'$ in constant time.
\end{theorem}

Combining \cref{thm:threshold-degreeminimal,thm:threshold-recognition,thm:dynamic-threshold-recognition} with \cref{obs:dynamic-edge-schemes,thm:general-algo}, we obtain a linear-time algorithm that decides whether a given weighted graph $(G,\omega)$ is a level-threshold graph. In the negative case, the algorithm of \cref{thm:dynamic-threshold-recognition} provides a certificate why the degree-minimal edge $e$ was not threshold-safe. Due to \cref{cor:theshold-certificate}, this forbidden induced subgraph is also contained in the level graph that is constructed by removing the remaining edges with the same weight as $e$. Thus, we can use this certificate to also prove that the respective level graph is not a threshold graph.

\begin{theorem}\label{thm:level-G-recognition-threshold}
  Let $\G$ be the class of threshold graphs.
  Then there exists a linear-time recognition algorithm for level-$\G$ weighted graphs.
  If the weighted graph is not level-$\G$, then the algorithm returns a level graph and a $2K_2$, $P_4$ or $C_4$ contained in it.
\end{theorem}

\subsection{Chain Graphs}\label{sec:chain}

We begin by characterizing the chain-safe edges.

\begin{lemma}\label{lemma:chain-safe}
  Let $G = (V,E)$ be a chain graph. An edge $e \in E$ is chain-safe if and
  only if $e$ is not the middle edge of an induced $P_4$ in $G$.
\end{lemma}

\begin{proof}
  By~\cref{thm:chain}, chain graphs are $2 K_2$-free. Therefore, the middle edge of an
  induced $P_4$ is not chain safe.  On the other hand, as the class of bipartite graphs is monotone, the deletion of every
  edge that is not a chain-safe edge must create a $2 K_2$ and,
  therefore, must be the middle edge of an induced $P_4$. 
\end{proof}

{We can now show that the prerequisites for Theorem 3.1 are all fulfilled.}

\begin{theorem}\label{thm:chain-degreeminimal}
  The class of chain graphs is degree sandwich monotone.
\end{theorem}

\begin{proof}
  {Let $G = (V,E)$ and $G' = (V,E\cup F)$ be chain graphs, where $E\cap F = \emptyset$.
    We need to show that every degree-minimal edge in $F$ is chain-safe.}
  Suppose for a contradiction that some degree-minimal edge in $F$, say
  $e = xy$, is not chain-safe. By \cref{lemma:chain-safe}, $e$ is the
  middle edge of an induced $P_4$, say $(u,x,y,v)$. Since the graph $G = G'-F$ is
  a chain graph and hence $2K_2$-free (see
  \cref{thm:chain}), at least one of the edges $ux$ and $vy$ belongs to
  $F$. Without loss of generality, we may assume that $ux\in F$. By the
  degree-minimality of $e = xy$ and \cref{lem:degree-minimal-neighbors},
  we have $d_{G'}(u)\ge d_{G'}(y)$. Since $v$ is adjacent to $y$ but not to
  $u$, there exists a vertex $t\in V(G')\setminus\{u,v,x,y\}$ such that $t$ is
  adjacent to $u$ but not to $y$. But now, the subgraph of $G'$ induced by
  $\{t,u,v,y\}$ is isomorphic to $2K_2$, a contradiction. 
\end{proof}

Similar to threshold graphs, the degree-minimality is needed here. For an example consider the graph from \cref{fig:counterexample} without the edge $wx$, which forms a chain graph.

The proof of \cref{thm:chain-degreeminimal} implies the same property as given for threshold graphs in \cref{cor:theshold-certificate}.

\begin{corollary}\label{cor:chain-certificate}
    Let $G$ be a chain graph and $F \subseteq E(G)$. If the degree-minimal edge $e$ in $F$ is not chain-safe, then every induced $2K_2$ in $G - e$ is also contained in $G - F$.
\end{corollary}%

\begin{theorem}[Heggernes and Kratsch~\cite{heggernes2007linear}]\label{thm:chain-recognition}
  There exists a linear-time recognition algorithm for chain graphs.
\end{theorem}

\begin{theorem}[{Beisegel et al.~\cite{MR4624523}}]\label{thm:chain-dynamic-recognition}
  There exists a dynamic recognition algorithm for chain graphs with linear preprocessing time and constant deletion time. If the graph $G'$ that results from the deletion of an edge is not a chain graph, then the algorithm returns an induced $2K_2$ of $G'$ in constant time.
\end{theorem}

Combining \cref{thm:chain-degreeminimal,thm:chain-recognition,thm:chain-dynamic-recognition} with \cref{obs:dynamic-edge-schemes,thm:general-algo} we obtain a linear-time algorithm that decides whether a given weighted graph $(G,\omega)$ is a level-chain graph.
Similarly as for threshold graphs, \cref{cor:chain-certificate} implies that this algorithm is certifying.

\begin{theorem}\label{thm:level-G-recognition-chain}
  Let $\G$ be the class of chain graphs.
  Then there exists a linear-time recognition algorithm for level-$\G$ weighted graphs. If the weighted graph is not level-$\G$, then the algorithm returns a level graph and a $2K_2$ contained in it.
\end{theorem}

\section{Conclusion}\label{sec:conclusion}

By combining the concepts of level graphs -- an important tool for the theory of Robinsonian matrices -- and edge elimination schemes, we give a sufficient condition for split-, \hbox{threshold-,} and chain-safe edges in order to generate so-called sorted safe-edge elimination schemes. This yields linear-time recognition algorithms for level-split, level-threshold, and level-chain weighted graphs.

The above-mentioned contributions raise some interesting questions. As the classes of chordal, chordal bipartite, and strongly chordal graphs are all sandwich monotone, it is natural to ask whether the weighted analogs of these classes can be recognized faster than checking every level graph separately. Note that we cannot adapt our approach directly since these graph classes are not degree sandwich monotone (see \cref{fig:chordal-counterexample}). For these classes, it seems to be difficult to quickly find a candidate for a $\G$-safe edge among the edges with minimal weight. 

It would be interesting to find similar results for graph classes which are not sandwich monotone, for example comparability graphs or interval graphs. Furthermore, it remains open whether weakly chordal graphs are sandwich monotone, a question also raised in \cite{heggernes2011strongly,sritharan2016graph}.

\begin{figure}
    \centering
    \begin{tikzpicture}[vertex/.style={inner sep=1.5pt,draw,circle}, scale=1.5]
    \tikzset{decoration={snake,amplitude=.4mm,segment length=2mm,%
          post length=0mm,pre length=0mm}}%
        \node[vertex, label=left: ] (1) at (0,0) {};
        \node[vertex, label=right: $u$] (2) at (1,0) {};
        \node[vertex, label=left: $v$] (3) at (0,1) {};
        \node[vertex, label=right: $w$] (4) at (1,1) {};
        \node[vertex, label=left: $ $] (5) at (0,1.5) {};
        \node[vertex, label=left: $ $] (6) at (0.75,1.5) {};
        \node[vertex, label=left: $ $] (7) at (1.25,1.5) {};

        \draw[thick] (4) -- (2) -- (1) -- (3);
        \draw[thick, decorate] (2) -- (3);
        \draw[thick, decorate] (3) -- (4);
        \draw[thick] (3) -- (5);
        \draw[thick] (4) -- (6);
        \draw[thick] (4) -- (7);

        \begin{scope}[xshift=4cm, yshift=0.25cm]
        \node[vertex, label=below: $w$] (1) at (0,0) {};
        \node[vertex, label=below: $u$] (2) at (1,0) {};
        \node[vertex, label=left: $ $] (3) at (2,0) {};
        \node[vertex, label=above: $x$] (4) at (0,1) {};
        \node[vertex, label=above: $v$] (5) at (1,1) {};
        \node[vertex, label=left: $ $] (6) at (2,1) {};
        \node[vertex, label=left: $ $] (7) at (-0.5,1.25) {};
        \node[vertex, label=left: $ $] (8) at (-0.5,0.75) {};
        \node[vertex, label=left: $ $] (9) at (-0.5,0.25) {};
        \node[vertex, label=left: $ $] (10) at (-0.5,-0.25) {};

        \draw[thick] (1) -- (2) -- (3) -- (6) -- (5) -- (4);
        \draw[thick, decorate] (2) -- (5);
        \draw[thick, decorate] (1) -- (4);
        \draw[thick] (1) -- (9);
        \draw[thick] (1) -- (10);
        \draw[thick] (4) -- (7);
        \draw[thick] (4) -- (8);
        \end{scope}
    \end{tikzpicture}
    \caption{The graph on the left side is strongly chordal and if we remove the wavy edge set $F$, then it is still strongly chordal. However, if we remove the unique degree-minimal edge $uv$ in $F$, then the graph is not chordal anymore, while removing the edge $vw$ is safe both for strongly chordal and chordal graphs. \\
    The graph on the right side is chordal bipartite and if we remove the wavy edge set $F$, then it is still chordal bipartite. Again, removing the unique degree-minimal edge $uv$ creates a graph that is not chordal bipartite, while removing edge $xw$ is safe.}
    \label{fig:chordal-counterexample}
\end{figure}

Our recognition algorithms do not work in linear time when arbitrary edge weights are considered. In this case, we first have to sort the edge weights using $\O(m \cdot \log n$) time. We leave open whether linear-time recognition algorithms exist also for arbitrary edge weights. Computing a sorted $\G$-safe edge elimination scheme does not seem to be a promising direction for such algorithms, at least not for split graphs. To see this, consider a $K_{1,m}$ with arbitrary edge weights. All level graphs of this graph are split graphs. So computing a sorted split-safe edge elimination scheme in linear time would imply that we can sort arbitrary weights in linear time.

Finally, let us mention two variations of the concepts discussed in this article that seem worthy of future investigations. In a natural extension, we can define and study the concepts of sandwich $k$-monotone graph classes for a positive integer $k$ by replacing the condition requiring the existence of a $\G$-safe edge in a particular set with the existence of a non-empty $\G$-safe subset of edges of cardinality at most $k$. Another variant of our problem arises when considering vertex weights instead. For graph classes that are not hereditary (for example, the connected graphs), one could examine their vertex-weighted analogs in which the level graphs are defined by deleting all sufficiently light vertices.

\section*{Acknowledgements}
The authors would like to thank Ulrik Brandes, Caroline Brosse, Christophe Crespelle, and Petr Golovach for their valuable discussions.

\bibliographystyle{plainurl}
\bibliography{weighted-graphs}

\end{document}